\documentclass[sigplan,10pt]{acmart}

\renewcommand\footnotetextcopyrightpermission[1]{} 
\pdfoutput=1

\usepackage[utf8]{inputenc}
\usepackage[T1]{fontenc}

\usepackage{graphics}
\usepackage{subfig}
\usepackage{tabu}
\usepackage{listings}
\usepackage{hyperref}
\usepackage{balance}

\usepackage[colorinlistoftodos]{todonotes}

\include{xcolor}
\definecolor{mygreen}{rgb}{0,0.6,0}
\definecolor{mygray}{rgb}{0.5,0.5,0.5}
\definecolor{mymauve}{rgb}{0.58,0,0.82}

\lstdefinestyle{lucscala}{
  language=Scala                   
}

\lstdefinestyle{lucoutput}{
  numbers=none,
  numbersep=0pt
}

\makeatletter
\renewcommand{\paragraph}{%
  \@startsection{paragraph}{4}%
  {\z@}{0.27ex \@plus 0.3ex \@minus 0.1ex}{-1em}%
  {\normalfont\normalsize\bfseries\itshape}%
}
\makeatother

\title{Tests as Maintainable Assets Via Auto-generated Spies}
\subtitle{A case study involving the Scala collections library's \texttt{Iterator} trait}

\author{Konstantin Läufer}
\orcid{0000-0002-7548-0876}
\affiliation{Loyola University Chicago}
\email{laufer@cs.luc.edu}

\author{John O'Sullivan}
\affiliation{Loyola University Chicago}
\email{josullivan@cs.luc.edu}

\author{George K.\ Thiruvathukal}
\orcid{0000-0002-0452-5571}
\affiliation{Loyola University Chicago and Argonne National Laboratory}
\email{gkt@cs.luc.edu, gkt@anl.gov}

\begin{document}

\newcommand{\listingcaptionskip}{0.5\baselineskip}
\newcommand{\listingsubcaptionskip}{-2.2\baselineskip}
\newcommand{\listingtopfigureskip}{-0.5\baselineskip}

\begin{abstract}
In testing stateful abstractions, it is often necessary to record interactions, such as method invocations, and express assertions over these interactions. 
Following the \emph{Test Spy} design pattern, we can reify such interactions programmatically through additional mutable state.
Alternatively, a mocking framework, such as Mockito, can \emph{automatically generate} test spies that allow us to record the interactions and express our expectations in a declarative domain-specific language.
According to our study of the test code for Scala's \path{Iterator} trait, the latter approach can lead to a significant reduction of test code complexity in terms of metrics such as code size (in some cases over 70\% smaller), cyclomatic complexity, and amount of additional mutable state required.
In this tools paper, we argue that the resulting test code is not only more maintainable, readable, and intentional, but also a better stylistic match for the Scala community than manually implemented, explicitly stateful test spies.
\end{abstract}


\maketitle

\newsavebox{\TempConvImperative}
\begin{lrbox}{\TempConvImperative}
\begin{lstlisting}[style=lucscala]
object TempConvImperative {
  var line = scala.io.StdIn.readLine()
  while (line != null) {
    val	raw = line.toInt
    val	celsius	= raw /	20
    println(celsius)
    line = scala.io.StdIn.readLine()
  }
}
\end{lstlisting}
\end{lrbox}

\newsavebox{\TempConvFunctional}
\begin{lrbox}{\TempConvFunctional}
\begin{lstlisting}[style=lucscala]
object TempConvFunctional extends App {
  val lines = scala.io.Source.stdin.getLines
  val values = lines.map(_.toDouble)
  val results = values.map(r => round(r / 20))
  results.foreach(println)
}
\end{lstlisting}
\end{lrbox}

\newsavebox{\TempConvOutput}
\begin{lrbox}{\TempConvOutput}
\begin{lstlisting}[style=lucoutput]
$ ./target/universal/stage/bin/temp-conv-imp
> 127
6
> 133
7
> 137
7
> 95
5
...
\end{lstlisting}
\end{lrbox}

\newsavebox{\CumAvgImperative}
\begin{lrbox}{\CumAvgImperative}
\begin{lstlisting}[style=lucscala]
object CumAvgImperative extends App {
  var count = 0
  var sum = 0.0
  var line = scala.io.StdIn.readLine()
  while (line != null) {
    count += 1
    sum += line.toDouble
    println(count + ": " + (sum / count))
    line = scala.io.StdIn.readLine()
  }
}
\end{lstlisting}
\end{lrbox}

\newsavebox{\CumAvgFunctional}
\begin{lrbox}{\CumAvgFunctional}
\begin{lstlisting}[style=lucscala]
object CumAvgFunctional extends App {
  val lines = scala.io.Source.stdin.getLines
  val values = lines.map(_.toDouble)
  val results = values.scanLeft((0, 0.0)) { 
    case ((count, sum), value) =>
      (count + 1, sum + value)
  }
  results.drop(1).foreach { case (count, sum) =>
    println(count + ": " + (sum / count))
  }
}
\end{lstlisting}
\end{lrbox}

\newsavebox{\CumAvgOutputCorrect}
\begin{lrbox}{\CumAvgOutputCorrect}
\begin{lstlisting}[style=lucoutput]
$ ./target/universal/stage/bin/cum-avg-imp
> 6
1: 6.0
> 7
2: 6.5
> 2
3: 5.0
> ^D
\end{lstlisting}
\end{lrbox}

\newsavebox{\CumAvgOutputIncorrect}
\begin{lrbox}{\CumAvgOutputIncorrect}
\begin{lstlisting}[style=lucoutput]
$ ./target/universal/stage/bin/cum-avg-fun
> 6
> 7
1: 6.0
> 2
2: 6.5
> ^D
3: 5.0
\end{lstlisting}
\end{lrbox}


\newsavebox{\scanLeftOrig}
\begin{lrbox}{\scanLeftOrig}
\begin{minipage}{\columnwidth}
\begin{lstlisting}[style=lucscala, firstnumber=595]
def scanLeft[B](z: B)(op: (B, A) => B): Iterator[B] = new Iterator[B] {
  var hasNext = true
  var elem = z
  def next() = if (hasNext) {
    val res = elem
    %*\bf\texttt{if (self.hasNext) elem = op(elem, self.next())}*)
    else hasNext = false
    res
  } else Iterator.empty.next()
}
\end{lstlisting}
\end{minipage}
\end{lrbox}

\newsavebox{\scanLeftTestSimple}
\begin{lrbox}{\scanLeftTestSimple}
\begin{minipage}{0.99\columnwidth}
\begin{lstlisting}[style=lucscala, firstnumber=9]
object Test {
  def main(args: Array[String]) {
    val it = Iterator.from(1).map(n => n * n).scanLeft(0)(_+_)

    assert(it.next == 0)
    assert(it.next == 1)
    assert(it.next == 5)
    // etc.
  }
}
\end{lstlisting}
\end{minipage}
\end{lrbox}


\newsavebox{\scanLeftFixed}
\begin{lrbox}{\scanLeftFixed}
\begin{lstlisting}[style=lucscala, firstnumber=604]
def scanLeft[B](z: B)(op: (B, A) => B): Iterator[B] = new AbstractIterator[B] {
  private[this] var state = 0    // 1 consumed initial, 2 self.hasNext, 3 done
  private[this] var accum = z
  private[this] def gen() = { val res = op(accum, self.next()) ; accum = res ; res }
  def hasNext = state match {
    case 0 | 2 => true
    case 3     => false
    case _     => if (self.hasNext) { state = 2 ; true } else { state = 3 ; false }
  }
  def next() = state match {
    case 0 => state = 1 ; accum
    case 1 => gen()
    case 2 => state = 1 ; gen()
    case 3 => Iterator.empty.next()
  }
}
\end{lstlisting}
\end{lrbox}


\newsavebox{\scanLeftTestOrig}
\begin{lrbox}{\scanLeftTestOrig}
\begin{lstlisting}[style=lucscala, firstnumber=300]
@Test def `scan is lazy enough`(): Unit = {
  val results = collection.mutable.ListBuffer.empty[Int]
  val it = new AbstractIterator[Int] {
    var cur = 1
    val max = 3
    override def hasNext = {
      results += -cur
      cur < max
    }
    override def next() = {
      val res = cur
      results += -res
      cur += 1
      res
    }
  }
  val xy = it.scanLeft(10)((sum, x) => {
    results += -(sum + x)
    sum + x
  })
  val scan = collection.mutable.ListBuffer.empty[Int]
  for (i <- xy) {
    scan += i
    results += i
  }
  assertSameElements(List(10,11,13), scan)
  assertSameElements(List(10,-1,-1,-11,11,-2,-2,-13,13,-3), results)
}
\end{lstlisting}
\end{lrbox}

\newsavebox{\scanLeftTestShortened}
\begin{lrbox}{\scanLeftTestShortened}
\begin{lstlisting}[style=lucscala, firstnumber=300]
@Test def `%*scan is lazy enough*)`(): Unit = {
  val results = ListBuffer.empty[Int]
  val it = new AbstractIterator[Int] {
    var cur = 1 ; val max = 3
    override def hasNext = {
      results += -cur ; cur < max }
    override def next() = {
      val res = cur ; results += -res
      cur += 1 ; res }
  }
  val xy = it.scanLeft(10)((sum, x) => {
    results += -(sum + x) ; sum + x
  })
  val scan = ListBuffer.empty[Int]
  for (i <- xy) { scan += i ; results += i }
  assertSameElements(List(10,11,13), scan)
  assertSameElements(List(10,-1,-1,-11,11,-2,-2,-13,13,-3), results) 
}
\end{lstlisting}
\end{lrbox}

\newsavebox{\scanLeftTestOrigOutput}
\begin{lrbox}{\scanLeftTestOrigOutput}
\begin{lstlisting}[style=lucoutput]
java.lang.AssertionError: expected:<List(10, -1, -1, -11, 11, -2, -2, -13, 13, -3)> but was:<ListBuffer(-1, -1, -11, 10, -2, -2, -13, ...)>
	at org.junit.Assert.fail(Assert.java:88)
	at AssertUtil.assertSameElements(...)
	at IteratorTest.scan_is_lazy_enough(IteratorTest.scala:316) %*\bf\texttt{<--~line~316~above}*) 
\end{lstlisting}
\end{lrbox}



\newsavebox{\scanLeftTestSpy}
\begin{lrbox}{\scanLeftTestSpy}
\begin{lstlisting}[style=lucscala, firstnumber=9]
  @Test def `%*scan is lazy enough with spy*)`() = {
    val it = spy(Iterator(1, 2, 3))
    val expected = Array(0, 1, 3, 6)
    val result = it.scanLeft(0)(_ + _)
    for (i <- expected.indices) {
      result.next() shouldBe expected(i)
      it.next() wasCalled i.times
    }
  }    
\end{lstlisting}
\end{lrbox}

\newsavebox{\scanLeftTestSpyOutput}
\begin{lrbox}{\scanLeftTestSpyOutput}
\begin{lstlisting}[style=lucoutput]
org.mockito.exceptions.verification.NeverWantedButInvoked:
elements.next();
Never wanted here:
-> at scala.collection.IndexedSeqLike$Elements.next(IndexedSeqLike.scala:59)
But invoked here:
-> at scala.collection.Iterator$$anon$14.next(Iterator.scala:600) %*\bf\texttt{<--~line~600~in~Figure~\ref{fig:scanLeftOrig}}*)

	at scala.collection.IndexedSeqLike$Elements.next(IndexedSeqLike.scala:59)
	at OurIteratorTest.$anonfun$scan_is_lazy_enough_with_spy$2(OurIteratorTest.scala:15) %*\bf\texttt{<--~line~15~above}*)
\end{lstlisting}
\end{lrbox}



\newsavebox{\testMethodsLOC}
\begin{lrbox}{\testMethodsLOC}
\begin{minipage}{\textwidth}
\begin{tabu}{|l|X|r|r|r|r|r|r|}
\hline
\bf{Method under test} 
& \bf{Test method} 
& \(\mathbf{Mut}_b\) 
& \(\mathbf{LOC}_b\) 
& \(\mathbf{LOC}_a\) 
& \(\mathbf{LOC}_\Delta\) 
& \(\mathbf{CC}_b\) 
& \(\mathbf{CC}_a\) 
\\ 
\hline
\path{Iterator.sliding} & \path{IteratorTest.}\texttt{grouped\-Iterator\-Should\-Not\-Ask\-For\-Unneeded\-Element} & 2 & 10 & 4 & -60\% & 3 & 1 \\
\hline
\texttt{Iterator.++} & \path{IteratorTest.}\texttt{no\-Excessive\-Has\-Next\-In\-Join\-Iterator} & 2 & 15 & 8 & -47\% & 6 & 2 \\
\hline
\texttt{Iterator.toStream} & 
\path{IteratorTest.}\texttt{to\-Stream\-Is\-Sufficiently\-Lazy} & 1 & 9 & 7 & -22\% & 3 & 3 \\
\hline
\path{Iterator.scanLeft} & \path{IteratorTest.}\texttt{\textasciigrave scan is lazy enough\textasciigrave} & 3 & 26 & 7 & -73\% & 5 & 3 \\
\hline
\path{HashMap.getOrElseUpdate} & \path{HashMapTest.}\texttt{get\-Or\-Else\-Update\_eval\-Once} & 1 & 5 & 4 & -20\% & 2 & 2 \\
\hline
\end{tabu}
\end{minipage}
\end{lrbox}




\noindent
\emph{Accepted for publication at the Tenth ACM SIGPLAN Scala Symposium (Scala ’19), July 17, 2019, London, United Kingdom}

\section{Introduction}
\label{sec:Introduction}

In our university-level programming languages course~\cite{Laufer:2018:lucproglangcourse}, the \texttt{scanLeft} method (similar to \texttt{foldLeft} with intermediate results) in the Scala collections library
is an important part of our overall pedagogy and the subject of many examples, including running averages and other forms of sliding analysis on unbounded streams, e.g., dynamic word clouds, stock market analysis, etc. 
In general, such prefix scans are useful and efficient building blocks for interactive, event-based, and other stream processing systems.

In this---and other---courses, we also emphasize the notion of \emph{tests as assets}. The software development community increasingly views automated tests as longer-term, maintainable assets along with the production code itself~\cite{10.1007/3-540-46020-9_35, Gonzalez:2017:LSU:3104188.3104236},
and a body of work on design patterns for automated testing has emerged~\cite{Meszaros:2006:XTP:1076526}. 
In particular, a \emph{Mock Object} replaces an object the system-under-test (SUT) depends on, is usually preconfigured to provide certain behaviors the SUT expects, and dynamically verifies the expected interactions coming from the SUT. 
By contrast, a \emph{Test Spy} also takes the place of a dependency of the SUT but behaves like the original dependency while recording the SUT's \emph{indirect outputs}, i.e., interactions with the dependency in terms of method invocation frequency and arguments, for later verification.
Both of these are subpatterns of \emph{Test Double}.


In the context of our course, we noticed and reported a seven-year-old bug in the \texttt{scanLeft} method of the \texttt{Iterator} trait, which provides some lazy stateful behaviors that are challenging to test. Indeed, the original test for \path{Iterator.scanLeft} does not fully test the correctness of this method under certain conditions.
This led us to study the Scala collections library's source code, where we noticed that the test suite includes several instances of manually implemented, explicitly stateful test spies.
While the corrected \texttt{scanLeft} implementation was successful in terms of clarity, conciseness, and idiomatic style, we found it difficult to understand the code and the actual cases being tested.

In this paper, we look for opportunities to use \emph{automatically generated} test spies as a systematic way to improve the tests for \texttt{scanLeft} and other \texttt{Iterator} methods and bring them in line with the notion of \emph{tests as assets}. 
While this serves as the underlying case study for this paper, the technique we describe is of general value as a \emph{programming pearl}. 
More broadly, it brings \emph{existing} tools and techniques to the Scala community in the hope that they will be useful. 
We have organized the rest of the paper as follows:
\begin{itemize}
    \item a detailed explanation of the case study based on the \path{scala.collection.Iterator} trait
    \item an overview of auto-generated test spies and how they can replace manually implemented test dependencies
    \item a side-by-side comparison of the official \texttt{scanLeft} test vs.\ an equivalent test with an auto-generated spy
\item a comparison of complexity metrics, including code size and cyclomatic complexity (a quantitative measure of the number of linearly independent paths through a program's source code)~\cite{McCabe:1976:CM:800253.807712}, before and after refactoring several tests from manually-implemented to auto-generated test spies, and 
    \item a summary of human and technical challenges that must be addressed for the proposed approach to be used more widely by Scala developers.
\end{itemize}


The novelty of our work is an initial exploration to identify, understand, and improve test assets, focused on a ``by hand'' analysis of the test code complexity. 
Although our efforts have been focused on the Scala library itself, owing to the critical role Scala's core library plays in all Scala development, the methods we describe are of general value and also play a role in our own development efforts. 
We are not aware of other work focused specifically on understanding test code complexity and using the resulting metrics to improve test assets. 
We believe our work has important implications in education and professional Scala development by training the next generation of Scala developers how to write effective, comprehensible, and maintainable tests.



\section{Case study: \texttt{Iterator.scanLeft}}

The \texttt{scala.collection.Iterator} trait did not have a \texttt{scan\-Left} method until it was requested in Scala Issue~4054~\cite{scalaissue4054} and implemented in January 2011 as shown in Figure~\ref{fig:scanLeftOrig}.

\begin{figure}
\captionsetup{skip=-0.8\baselineskip}
\usebox{\CumAvgFunctional}
\caption[Cum Avg]{
The imperative version of the cumulative running average filter is a simple, monolithic while loop. 
The functional version shown here is a pipeline of modular, separately testable stages, arguably making it more readable and maintainable.
Both run in linear time and constant space. 
}
\label{fig:CumAvgCode}
\end{figure}

\begin{figure*}
\vspace*{\listingtopfigureskip}
\captionsetup{skip=\listingcaptionskip}
\captionsetup[subfloat]{captionskip=\listingsubcaptionskip}
\subfloat[correct behavior]{
\usebox{\CumAvgOutputCorrect}
\label{fig:CumAvgOutputCorrect}
}
\hspace{\columnsep}
\subfloat[incorrect behavior]{
\usebox{\CumAvgOutputIncorrect}
\label{fig:CumAvgOutputIncorrect}
}
\caption{Correct (left) and incorrect (right) sample runs of the cumulative running average filters. 
In the incorrect case, the updated average goes up to the previous instead of the current input value. 
(We prefix input lines with \texttt{>}.)
}
\label{fig:CumAvgOutput}
\end{figure*}

\begin{figure*}
\vspace*{\listingtopfigureskip}
\captionsetup{skip=-\baselineskip}
\subfloat{\usebox{\scanLeftOrig}}
\hspace{\columnsep}
\subfloat{\usebox{\scanLeftTestSimple}}
\caption{Original version of \texttt{scanLeft} and associated correctness test.
In this implementation, the resulting iterator does not return the current item until \emph{after} the (premature) call to \texttt{self.next()} on line 600 returns (shown in boldface above).
The test does not catch this bug because it focuses on the overall correctness of the sequence of items returned.
}
\label{fig:scanLeftOrig}
\end{figure*}

\begin{figure*}
\vspace*{\listingtopfigureskip}
\captionsetup{skip=\listingcaptionskip}
\captionsetup[subfloat]{captionskip=-\baselineskip}
\subfloat[test with manually implemented spy]{
\begin{minipage}{\columnwidth}
\centering
\usebox{\scanLeftTestShortened}
\\
\vspace*{-\baselineskip}
\usebox{\scanLeftTestOrigOutput}
\end{minipage}
\hspace{\columnsep}
\label{fig:scanLeftTestManual}
}
\subfloat[test with auto-generated spy]{
\begin{minipage}{0.95\columnwidth}
\centering
\usebox{\scanLeftTestSpy}
\\
\vspace*{-\baselineskip}
\usebox{\scanLeftTestSpyOutput}
\end{minipage}
\label{fig:scanLeftTestSpy}
}
\caption[Incremental Tests]{Incremental correctness and laziness tests for \texttt{Iterator.scanLeft}. (a) Manually implemented, explicitly stateful spy defined (interwoven with the SUT itself) on lines 301--309 and 311, exercised on line 314, and verified on line 316.
(b) Auto-generated, declarative spy using Mockito Scala defined on line 10, exercised on lines 13--14, and verified on line 15.
We argue that the failure message for (b) is more helpful by directly indicating the site of the unwanted invocation of \texttt{next()}.
}
\label{fig:scanLeftTests}
\end{figure*}



This version, however, causes the example from Figure~\ref{fig:CumAvgCode} to behave incorrectly as seen in Figure~\ref{fig:CumAvgOutputIncorrect}: Instead of printing the first updated average right after reading the first value, it prints this only after reading the second value; 
it then prints each subsequent update delayed by one input value, and the final update only \emph{after} EOF.
The reason is that the iterator returned by \texttt{scan\-Left} does not return the current item until \emph{after} the (premature) call to \texttt{self.next()} on line 600 in Figure~\ref{fig:scanLeftOrig} returns.
The included test does not catch this bug because it focuses on the correctness of the resulting items, irrespective of the interactions with the original iterator.



We reported this bug as Scala Issue~10709~\cite{scalaissue10709} in February 2018, after attempting to use \texttt{scanLeft} in the context of our spring 2018 Scala-based programming languages course~\cite{Laufer:2018:lucproglangcourse}. 
The Scala team promptly fixed this issue as of Scala 2.12.5, by reimplementing the method using a flat four-state machine, replaced as of Scala 2.13.x~\cite{scala:collection-strawman} with an arguably more elegant and straightforward implementation based on the State pattern~\cite{Gamma:1995:DPE:186897}.

The corresponding test, shown in Figure~\ref{fig:scanLeftTestManual}, uses a test spy in the form of a custom iterator with additional state to test for incremental correctness along with the ``right amount of laziness.''



\begin{figure}
    \includegraphics[width=\columnwidth]{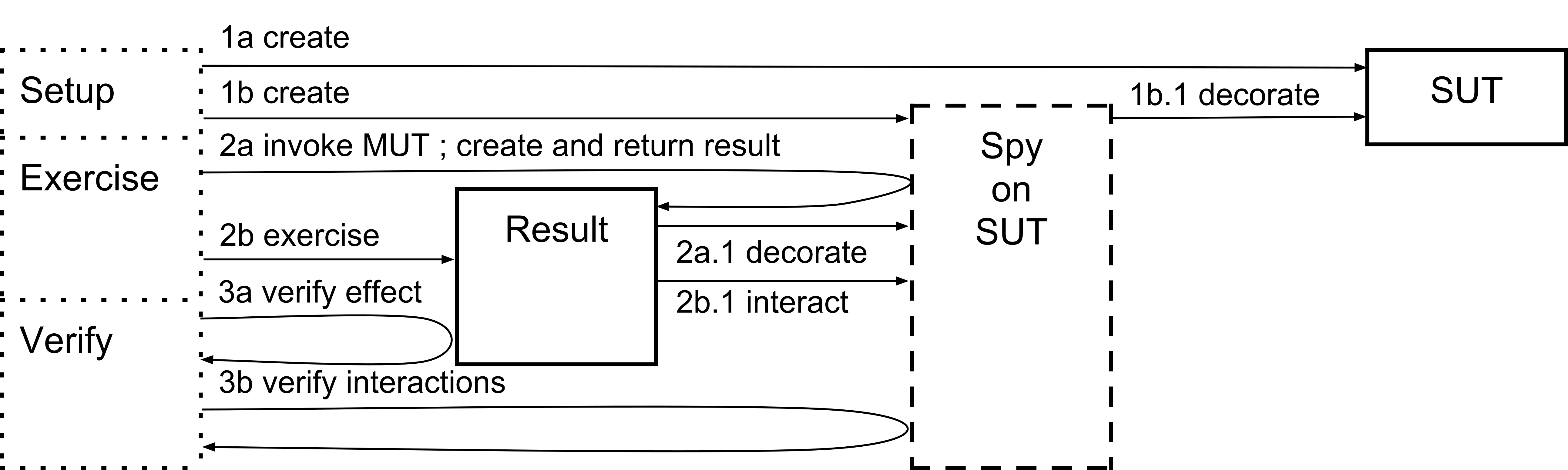}
    \vspace*{-1.5\baselineskip}
    \caption{Collaboration diagram for a spy-based test of a stateful SUT's method that returns a decorator of the SUT. The test verifies through the spy that the interactions with the MUT's result have the desired effect on the original SUT.}
    \label{fig:TestWithSpy}
\end{figure}

\paragraph{Discussion}
There are several possible factors contributing to the fact that this bug remained unreported for so long:
\begin{itemize}
\item \texttt{Iterator.scanLeft} might be rarely used, or rarely used \emph{incrementally} as described above.
\item Developers might have resorted to a workaround but not taken the time to report the actual bug.
\item State-dependent behaviors are challenging to comprehend, document, implement, and test.
\end{itemize}



Furthermore, there is considerable \emph{essential complexity}~\cite{Brooks:1987:NSB:26440.26441} to a stateful behavior such as \texttt{scan\-Left}: it returns a decorator~\cite{Gamma:1995:DPE:186897} around the receiver, after which one is no longer allowed to interact directly with the receiver, while subsequent interactions with the decorator have side effects on the original receiver, such as reading lines from input.
This complexity carries over to the spy-based test in terms of the dynamic interactions shown in Figure~\ref{fig:TestWithSpy}.


We conjecture that tests that rely on manually implemented spies typically suffer from two additional shortcomings: 
(1) \emph{accidental complexity} that affects maintainability, and (2) unhelpful failure messages.
We will now explore how automatically generated spies can help address both issues.


\begin{table*}
\usebox{\testMethodsLOC}
\vspace*{0.5\baselineskip}
\caption{Comparison of test code complexity metrics (number of mutable objects, LOC, and Cyclomatic Complexity counted manually) before/after refactoring from manually implemented to auto-generated test spies. 
(\(\mathbf{Mut}_a = 0\) for all test methods.)
Even when there is no or little reduction in the metrics, there is usually a gain in readability and clarity of intent.
This table includes most affected tests in \texttt{IteratorTest} and \texttt{HashMapTest}; there may be other applicable tests elsewhere in the codebase.
}
\label{tab:codesize}
\end{table*}

\section{Auto-generated test spies can help}


As an alternative to manually implementing test doubles (see Section~\ref{sec:Introduction}) to represent the SUT's dependencies, mocking frameworks support the automatic generation of mock objects, usually based on their type.
Some mocking frameworks, such as Mockito~\cite{mockito}, additionally support the automatic generation of \emph{test spies}.
We chose Mockito because of its maturity, support for idiomatic Scala syntax~\cite{mockitoscala}, and unique support for spying on final and anonymous classes.

Concretely, as shown in Figure~\ref{fig:scanLeftTestSpy}, we can use Mockito to wrap a test spy around a simple iterator instance, invoke \texttt{scanLeft}, and then interact with the iterator resulting from this invocation;
these interactions still correspond to Figure~\ref{fig:TestWithSpy}.
During these interactions, we can test not only the (overall) correctness of the values of the resulting iterator, but also the (incremental) correctness of the effects of these interactions on the original iterator. 
In this case, the correct amount of laziness means to have invoked \texttt{next()} on the original iterator \texttt{i} times, where \texttt{i} is the position of the current value in the resulting iterator.
Specifically, when we invoke \texttt{next()} on the resulting iterator for the first time, we expect to see the initial \texttt{z} value of \texttt{scanLeft}, and there should not yet have been any invocations of \texttt{next()} on the original iterator.

This test is three times as short as the official version. 
We argue that it is not only more comprehensible, maintainable, and effective at conveying the intent, but it also produces a more useful error message pointing directly to the offending eager invocation of \texttt{next()}.

\paragraph{Discussion}
    By eliminating the need for custom iterator implementations with mutable state, auto-generated test spies allow the programmer to focus on the functional correctness of the SUT.
This promotes the view of test code as readable, comprehensible, teachable, and maintainable assets, which has the potential to work its way back into to the API documentation.
If more developers feel encouraged to write tests, this might foster a test-driven mindset in the community.

Table~\ref{tab:codesize} shows other test methods that can benefit from this approach;
those typically exhibit \emph{code smells} such as \texttt{var} or mutable state other than the stateful SUT, e.g., a buffer. 
Because Test Spy is a common subpattern of Test Double~\cite{Meszaros:2006:XTP:1076526}, we expect the technique of spying on a stateful SUT to be beneficial in similar scenarios to Figure~\ref{fig:TestWithSpy}, even when the result of the method-under-test (MUT) depends on the SUT in more general ways than decorating/wrapping the SUT.
For instance, we have used generated test spies for unit testing a view component in a Scala-based Android app~\cite{Laufer:2014:stopwatch-android-scala}.





\section{Conclusions and Future Work}


We have shown that Mockito's auto-generated test spies eliminate much accidental complexity from certain state-based tests.
We plan to conduct a similar investigation of Scala \texttt{view}, \texttt{Stream}, and the \texttt{LazyList} class added in 2.13.x.

More broadly, we hope to use repository mining to identify other Scala projects that can benefit from making tests more immutable and declarative, understand the effect of these refactorings on test code complexity/quality~\cite{Bass:2012:SAP:2392670} and process metrics~\cite{Thiruvathukal:metricsdashboard}, and investigate the possibility of tool support.

Finally, it turns out that a test coverage tool would not have indicated the problem with the original test for \path{Iterator.scanLeft} shown in Figure~\ref{fig:scanLeftOrig}. 
Common metrics, such as statement and branch coverage, remain unchanged (after compensating for a finite vs.\ indefinite iterator as SUT). 
Further study might reveal whether suitable existing test coverage metrics are effective for these complex stateful behaviors.
Nevertheless, taming test code complexity using test spies can be an effective strategy for improving the comprehension and maintainability of test cases.


\paragraph{Acknowlegments}
We are grateful to the Scala team for their responsiveness to our bug report; and to Bruno Bonanno for suggesting we use  Mockito Scala in our examples.

\bibliographystyle{ACM-Reference-Format}
\bibliography{references}
\balance

\end{document}